\documentclass[graybox]{svmult}

\usepackage{mathptmx}       
\usepackage{helvet}         
\usepackage{courier}        
\usepackage{type1cm}        
\usepackage{makeidx}         
\usepackage{graphicx}        
\usepackage{multicol}        
\usepackage[bottom]{footmisc}

\makeindex             

\begin{document}

\title*{Uncovering hidden modes in RR Lyrae stars}
\author{L. Moln\'ar, Z. Koll\'ath and R. Szab\'o}
\institute{L. Moln\'ar, Z. Koll\'ath, R. Szab\'o \at Konkoly Observatory, MTA CSFK, Budapest, 1121, Konkoly Thege Mikl\'os \'ut 13-17. \email{lmolnar@konkoly.hu}}
%
%
\maketitle

\abstract*{The Kepler space telescope revealed new, unexpected phenomena in RR Lyrae stars: period doubling and the possible presence of additional modes. Identifying these modes is complicated because they blend in the rich features of the Fourier-spectrum. Our hydrodynamic calculations uncovered that a 'hidden' mode, the 9th overtone is involved in the period doubling phenomenon. The period of the overtone changes by up to 10 per cent compared to the linear value, indicating a very significant nonlinear period shift caused by its resonance with the fundamental mode. The observations also revealed weak peaks that may correspond to the first or second overtones. These additional modes are often coupled with period doubling. We investigated the possibilities and occurrences of mutual resonances between the fundamental mode and multiple overtones in our models. These theoretical findings can help interpreting the origin and nature of the 'hidden' modes may be found in the high quality light curves of space observatories.}

\abstract{The Kepler space telescope revealed new, unexpected phenomena in RR Lyrae stars: period doubling and the possible presence of additional modes. Identifying these modes is complicated because they blend in the rich features of the Fourier-spectrum. Our hydrodynamic calculations uncovered that a 'hidden' mode, the 9th overtone is involved in the period doubling phenomenon. The period of the overtone changes by up to 10 per cent compared to the linear value, indicating a very significant nonlinear period shift caused by its resonance with the fundamental mode. The observations also revealed weak peaks that may correspond to the first or second overtones. These additional modes are often coupled with period doubling. We investigated the possibilities and occurrences of mutual resonances between the fundamental mode and multiple overtones in our models. These theoretical findings can help interpreting the origin and nature of the 'hidden' modes may be found in the high quality light curves of space observatories.}

\section{Introduction}
\label{intro}
Before the advent of space-based, quasi-continuous observations, the light variation of RR Lyrae stars were considered to be relatively simple: they either pulsate in fundamental mode or in first overtone, or in both in the case of double-mode stars. The only complicating factor has been the Blazhko-effect, the periodic modulation of the amplitude and the phase of the pulsation. One particular model proposed that it is caused by a resonance between the fundamental mode and a large-amplitude non-radial mode (\cite{dc99}, \cite{nd01}), but the resonance model was not confirmed by observations yet. Resonances that may excite overtones and cause period doubling were also sought but were not found between the fundamental mode and the first four overtones in a model survey (\cite{mb90}). 

\section{Period doubling}
\label{pd}
However, period doubling, the alternation of pulsation cycles with smaller and larger amplitudes was unexpectedly found in the first observations of RR Lyr itself by the \textit{Kepler} space telescope (\cite{rrlfirst}). Eventually two other stars were found where the alternations were clearly visible in the data and four were the spectrum displayed half-integer frequency peaks ($f = n/2 f_0$), the spectral features of period doubling, bringing the total count to seven (\cite{szabo10}, \cite{benko10}). The phenomenon was detected in modulated stars only.

Serendipitously, period doubling was also encountered in hydrodynamic models not long before the \textit{Kepler} observations. After an extensive model survey, we were able to show that the phenomenon is caused by a 9:2 resonance between the fundamental mode and the 9th overtone (\cite{kmsz11}). The high order overtone was found to be a strange mode. The resonance displays surprising strength: while the period of the pulsation cycle is shifted by less than 0.5 percent compared to the linear period value of the fundamental mode, the period of the 9th overtone can be changed by as much as 5 to 10 percent. 

\begin{figure}[b]
\sidecaption
\includegraphics[scale=.45]{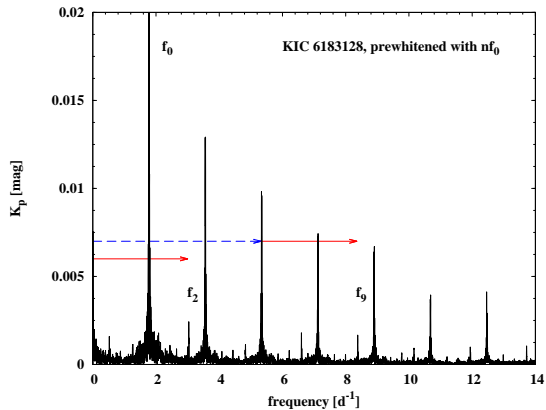}
\caption{Fourier-spectrum of \textit{Kepler} Q1-Q2 data of V354 Lyr (KIC 6183128), prewhitened with the pulsation frequency and it's harmonics. Note the higher amplitude $nf_0+f_2$ peaks at 6.6 and 8.4 c/d. The latter may indicate the presence of a resonant $f_9$ mode. Solid and dashed arrows visualise the possible $3 f_0 +f_2 = f_9$ resonance.}
\label{sp}       
\end{figure}

\section{Additional modes}
Significant peaks besides the pulsation frequency, it's harmonics and the modulation sidelobes were detected in RR Lyrae Fourier spectra both with the CoRoT (\cite{v1127},\cite{poretti},\cite{gugg}) and \textit{Kepler} space telescopes (\cite{benko10}). They appear close to the period ratios of the first or second overtones, but the amplitude difference compared to the fundamental mode is so large that such stars cannot be considered as classical double mode pulsators. Out of the six stars with similar additional peaks, four show both the Blazhko-effect and period doubling too. Two stars however (KIC 7021124 and 9508655) display only marginal signs of modulation with sidelobe amplitudes below 1.5 and 0.5 mmag respecively, that can be attributed to the residual amplitude differences between the different quarters of the data as well (\cite{benko10}, \cite{nemec}).

\begin{figure}[b]
\sidecaption
\includegraphics[scale=.37]{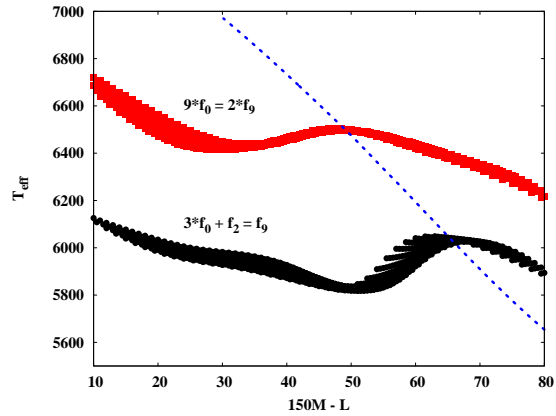}
\caption{The diagnostic diagram with two resonance regions included. Areas covered with rectangles and points represent linear models where $9f_0 = 2f_9$ and $3f_0 + f_2 = f_9$, respectively. The former is responsible for the period doubling. The dashed line shows the linear models fitted to the $f_0$ and $f_2$ frequency values of V354 Lyr (KIC 6183128). }
\label{3res}       
\end{figure}

An interesting case is V354 Lyr (KIC 6183128), where the $nf_0+f_2$ combination peaks do not decrease monotonically, as expected, suggesting some resonance with an undetected mode (Figure \ref{sp}). The increase occurs in the frequency region of the 9th overtone, the strange mode. Our current hypothesis is a three-mode resonance between the fundamental mode and the two overtones, in the from of $3f_0 + f_2 = f_9$. The position of the linear three-mode resonance region in our diagnostic diagram ($T_{eff}$ vs. $150\,M-L$, see \cite{kmsz11}) is diplayed in Figure \ref{3res}.

The period ratios of the fundamental and additional modes can be used to determine the physical characteristics of these stars. We could not find nonlinear models with a three-mode resonance and excited additional modes yet, however the strong period shifts we found in the 9:2 resonance case indicate that resonant nonlinear models may be located at a considerably different parameter range compared to the intersection of the linear results, displayed in Figure \ref{3res}.

\section{Future prospects}
Radial resonances gained more interest after the emergence of a new Blazhko-model. The 9:2 resonance, incorporated in amplitude equations, produced not only period doubling but also amplitude modulation (\cite{bk11}). It is unclear however, how far are the modulated AE solutions from the hydrodynamic models of RR Lyrae stars. Finding more resonances and fitting the corresponding coefficients of AEs from hydrodynamic calculations might help such investigations.

\textit{Kepler} revealed a whole variety of additional frequency peaks in RR Lyrae stars around and below the millimagnitude level. Half of the modulated stars in the \textit{Kepler} RRab sample display either period doubling or other signs or both. Additional peaks are not limited to the first or second overtone candidates but are found at different values as well, suggesting the presence of additional resonances and the possibility of nonradial modes too. 

From the theoretical side, we also detected peculiar hydrodynamic solutions with irregularly varying amplitudes that were found to be chaotic solutions (\cite{plachy}). The \textit{Kepler} long cadence observations hinted some irregularity in period doubling that might be the manifestation of this chaotic behaviour but short cadence observations and careful analysis will be needed to settle this issue.

\begin{acknowledgement}
This work was supported by the Hungarian OTKA grants K83790 and MB08C 81013. RSz acknowledges the Bolyai J\'anos Scholarship of the Hungarian Academy of Sciences. The research leading to these results has received funding from the European Community's Seventh Framework Programme (FP7/2007-2013) under grant agreement no. 269194. Funding for the \textit{Kepler} mission is provided by the NASA Science Mission directorate.
\end{acknowledgement}

\end{document}